\documentclass[a4paper]{article}
\usepackage{amssymb,amsmath}
\usepackage{graphics,epsf}
\usepackage{floatflt}

\renewcommand{\Re}{\hbox{Re}\hspace{.2pt}}
\begin{document}
\voffset=-30mm \hoffset=-20mm \setcounter{page}{734} JOURNAL OF
APPLIED SPECTROSCOPY\hfill Volume 12, Number 6\hfill June 1970\\
\copyright 1973 {\it Consultants Bureau, a division of Plenum Publishing Corporation,
227 West 17th Street, New York, N.Y. 10011. All rights reserved. March 1, 1973.
Translated from Zhurnal Prikladnoi Spektroskopii, Vol. 12, No.6, pp. 989-993, June,
1970. Original article submitted May 5, 1969}.\vspace{5mm}\hrule\vspace{20mm}

\noindent EFFECT  OF RESONANCE  RADIATIVE PROCESSES\\
ON THE  AMPLIFICATI0N
FACTOR \vspace{5mm}

T. Ya. Popova and A. K. Popov \hfill UDC 541.15 \vspace{10mm}

The effect of a strong held in resonance with one of the
transitions of a substance is to change the emission and
absorption spectra of adjacent transitions [1-3]. These changes
are due to three effects:  variation of the common-level
population, splitting of energy levels in a strong field, and
nonlinear interference processes. The last effect can be
interpreted as an interference and a modification of the
frequency-correlation properties of step-by-step and two-photon
processes in a strong field under resonance conditions [4, 5].
Undoubtedly, these effects should manifest themselves in the
amplification (absorption) properties of a weak held at one
transition in the presence of a strong field in resonance with an
adjacent transition. Of particular interest are optical
transitions which have a greater number of relaxation constants
than microwave transitions. Any of the above-mentioned effects can
be made to predominate by appropriate selection of transitions.
Maximum change of the amplification (absorption) factor can be
obtained in gases for uniformly broadened transitions. {\it It will
be shown below that amplification can occur even it the lower level
has a greater saturated population than the upper.}

We shall analyze the case when both fields are in resonance, as
the change in amplification is then the greatest. To be specific,
consider the scheme of transition with a common lower level shown
in Fig. 1.  The system is located in a strong field whose
frequency $\omega$ is equal to the natural transition frequency
$\omega_{mn}$ ($\omega=\omega_{mn}$). The amplification factor of
a weak field at the center of the transition line $gn$ can be
found from $$
\alpha_\mu=-\hbar\omega_{gn}\left[c|E_\mu|^2/8\pi\right]^{-1}N
2\Re\{iV_{gn}e^{i\omega_{gn}t}\rho_{ng}\}. $$ Here $E_{\mu}$ is
the amplitude of the weak field at frequency $\omega_{mu}
=\omega_{gn} $, $N$ is the number of particles per unit volume,
$V_{gn}$ is a matrix element of the perturbation Hamiltonian for
the weak field $V_{gn}=G_\mu \exp\{-i (\omega_{gn}t-{\bf k}_\mu
\mathbf{r})\}$; $G_\mu=-\mathbf{E}_\mu \mathbf{d}_{gn}/2\hbar$,
$\rho_{gn}$ is a density matrix in the interaction representation.
The system of equations for the density matrix has a standard form
and becomes algebraic in the stationary case [3]. The particle
velocity distribution considered in [3] was Maxwellian.
\begin{figure}[!h]
\center{\epsfxsize=.47\textwidth\leavevmode\epsfbox{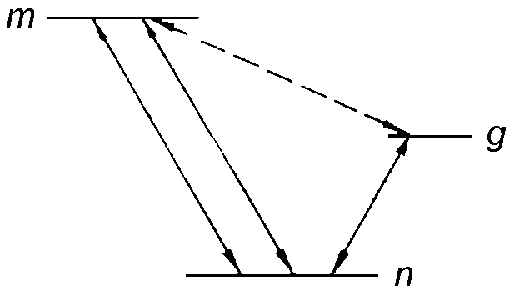}\hspace{12pt}
\epsfxsize=.35\textwidth\leavevmode\epsfbox{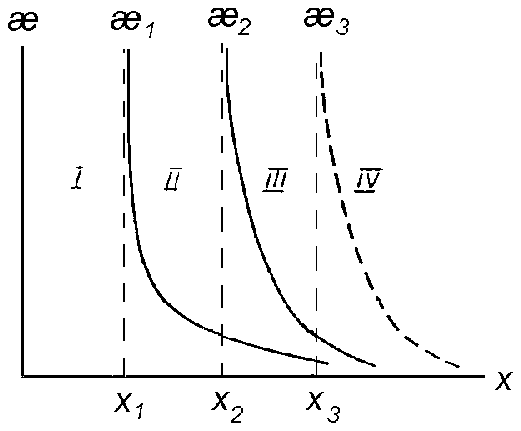}}
\parbox[t]{.47\textwidth}{\caption{}}
\hfill
\parbox[t]{.47\textwidth}{\caption{}}
\end{figure}

Here we consider a monoenergetic particle beam with an arbitrary
velocity $v$ (this includes stationary particles).  The change in
amplification factor can be the greatest in such a case; on the
other hand, this example makes it possible to find the interaction
of the field with an individual atom. The monoenergetic beam can
be real and effective. An effective beam can be obtained by
exciting atoms with a secondary monochromatic field onto one of
the higher levels and letting them relax to the considered levels.
A monochromatic field in a gas with nonuniform spectral line
broadening interacts only with atoms the velocity projection of
which in the $\Delta v_0\approx 2\gamma_0/k_0$ interval is near
the "resonance" velocity with a projection $v_0$ in the
$\mathbf{k}_0$ direction.  Here $k_0$ is the absolute value of the
wave vector of the secondary excitation field, and $2\gamma_0$ is
the line-width for an individual atom at the excitation
transition. It is thus possible to ensure that the atoms have a
negligible velocity spread on levels with which the nonlinear
interference processes are associated.  To this case also belong
transitions for which the sum of the collision and natural widths
is comparable with the Doppler width, for example, the $3s_2-3p_4$
transition in neon.

In a monoenergetic beam $\omega_{gn}$ and $\omega_{mn}$ should be
understood as the Doppler-shifted natural frequencies of the
corresponding transitions. In this case, the amplification factor
$\alpha_\mu$ at the center of the line and the population
difference $N (n_g-\rho_{nn})$ saturated by the strong field are
respectively
\begin{eqnarray}
&n_g-\rho_{nn}=\Delta n_{gn}-\Delta
n_{mn}\left(1-\dfrac{\gamma_{mn}}{\Gamma_m}
\right)\dfrac{2\Gamma}{\Gamma_n}\dfrac{|G|^2}{\Gamma^2(1+\varkappa)},&\\
&\dfrac{\alpha_\mu}{\alpha_\mu^0}=\dfrac{\Gamma_{gn}}{\Gamma_{gn}+|G|^2
\Gamma_{gm}^{-1}}\left\{1-\dfrac{\Delta n_{mn}}{\Delta
n_{gn}}\dfrac{|G|^2} {\Gamma^2(1+\varkappa)}
\left[\left(1-\dfrac{\gamma_{mn}}{\Gamma_m}\right)\dfrac{2\Gamma}{\Gamma_n}+
\dfrac{\Gamma}{\Gamma_{gm}}\right]\right\}.&
\end{eqnarray}
\noindent Here $N\Delta n_{ik}=N(n_i-n_k)$ are unsaturated
population differences at the respective transitions (for $|G|^2 =
0$); $N$ is the particle concentration at the ground state; $G$ is
a matrix element of the perturbation Hamiltonian of the strong
field: $G =-{\bf Ed_{mn}}/2\hbar$; $\hbar\Gamma_i$ are the energy
level widths; $\Gamma_{ik}$ $(\Gamma_{mn}=\Gamma)$ usually appear
as half- widths of the corresponding transition lines (in angular
frequencies); and $\gamma_{mn}$ is the probability of relaxation
transition from level $m$ to lever $n$ per unit time. The
saturation parameter $\varkappa$ is given by $$
\varkappa=(\Gamma_m+\Gamma_n-\gamma_{mn})(\Gamma_m\Gamma_n\Gamma)^{-1}2|G|^2=
\tau^2 2|G|^2, $$ and $\alpha_\mu^0$ is the amplification factor
at the line center without the strong field. The term proportional
to $|G|^2$ in the denominator of the first factor in (2) reflects
the effect of level splitting in the strong field, while the term
inversely proportional to $\Gamma_{gm}$ accounts for the presence
of frequencv-correlated transitions between the states $m$ and $g$
via an intermediate level (nonlinear interference processes) [4,
5].

From (1) follows that an inversion of the sign of $n_g-\rho_{nn}$
should take place if the signs of $\Delta n_{gn}$ and $\Delta
n_{mn}$ are the same and if the field obeys the relation
\begin{equation}
\left(1-\dfrac{\gamma_{mn}}{\Gamma_{m}}\right)\dfrac{2|G|^2/\Gamma\Gamma_n}{1+
\varkappa}>\dfrac{\Delta n_{gn}}{\Delta n_{mn}}
\end{equation}
{\it However, the contribution of interference effects causes
$\alpha_\mu$ sign inversion to take place at lower field values}
provided the inequality
\begin{equation}
\left[1-\dfrac{\gamma_{mn}}{\Gamma_m}+\dfrac{\Gamma_{n}}{2\Gamma_{gm}}\right]
\dfrac{2|G|^2/\Gamma\Gamma_n}{1+\varkappa}>\dfrac{\Delta
n_{gn}}{\Delta n_{mn}}
\end{equation}
is satisfied.

Comparing (3) and (4) we observe that the difference between the critical
fields in the first and second case is the greater the stronger the
inequality
\begin{equation}
\Gamma_{gm}\leq\Gamma_n/2\left(1-\dfrac{\gamma_{mn}}{\Gamma_m}\right).
\end{equation}
In particular, if $\Gamma_m=\gamma_{mn}$ the external field has no
effect on the population of level $n$ and the change of the sign
of $\alpha_\mu$ is due solely to interference effects. If
inequality (3) is observed and (5) is not, the absolute value of
the inverted population difference $|n_g-\rho_{nn} (\varkappa)|$
will for some fields exceed the interference term $\Delta
n_{mn}|G|^2/\Gamma_{gm}\Gamma(1+\varkappa)$ and cancel its effect.

Figure 2 shows schematically the analysis of the conditions of
inversion of $\alpha_\mu$ sign at the line center provided $\Delta
n_{nm}$ and $\Delta n_{gn}$ have the same signs.  The values of
$\varkappa$ and $x = \Delta n_{mn}/\Delta n_{gn}$ are plotted along
the $Y$ and $X$ axes respectively. The curve $\varkappa_1(x)$
describes fields for which $\alpha_{\mu}$  turns into zero. In
domain I no sign inversion can take place whatever the field. {\it
In domain II sign inversion takes place without a change of the
sign of the population difference $n_g-\rho_{nn}(\varkappa)$}. The
curve $\varkappa_2 (x)$ describes fields for which this difference
is zero. In domain III the population difference
$n_g-\rho_{nn}(\varkappa)$ changes its sign. In domain IV the
interference contribution is counterbalanced by the population
difference in case condition (5) is not satisfied.

\begin{floatingfigure}{60mm}
\epsfxsize=50mm \center{\leavevmode\epsfbox{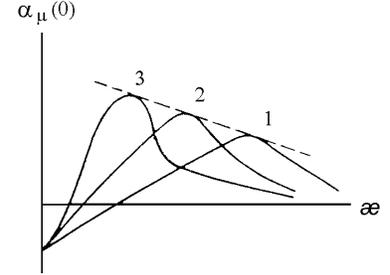}}
\vspace{-5mm}\caption{ $\alpha_\mu(0)$ as a function of $\varkappa$.
$x=\Delta n_{mn}/\Delta n_{gn}>0$ increases from curve $1$ to curve $3$ ($
x^{(1)}<x^{(2)}<x^{(3)}$).  }
\end{floatingfigure}

The expressions for curves $\varkappa_1$, $\varkappa_2$,
$\varkappa_3$ and for the critical population difference ratios are
\begin{eqnarray}
\varkappa_i(x)=(xx_i^{-1}-1)^{-1};\quad
x_2=\dfrac{\Gamma_m-\gamma_{mn}+\Gamma_n} {\Gamma_m-\gamma_{mn}};\nonumber\\
x_{1,3}=\dfrac{\Gamma_m-\gamma_{mn}+\Gamma_n}
{\Gamma_m-\gamma_{mn}\pm \Gamma_n\Gamma_m(2\Gamma_{gm})^{-1}}.
\label{6}
\end{eqnarray}
{\it The fact that amplification is possible when $x < x_2$,
$\varkappa<\varkappa_2$, even if $n_g-\rho_{nn}(\varkappa)<0$} can
be explained as follows. For relatively low external fields the
nonlinear interference effects can be considered as a contribution
of two-photon transitions $g\leftrightarrow m$ via an intermediate
level $n$ to the emission (absorption) of $\hbar\omega_\mu$ quanta
(4). For amplification to take place at the frequency $\omega_\mu$
it is sufficient that $n_g>\rho_{mm}$ irrespective of the sign of
$n_g-\rho_{nn}$ (so that always $x_i>1$).  The range of $x$
values, $x_2$ to $x_1$, for which {\it inversion of the
$\alpha_\mu$ sign can take place without inversion of
$n_g-\rho_{nn}$} is the greater the stronger the inequality (5).

In solids, in which linewidths are as a rule considerably broader than
levels, condition (5) can be satisfied only when $\Gamma_m\approx \gamma_{mn}$.
If this condition is satisfied, the strong and weak fields interact mainly
through interference processes and
\begin{equation}
x_1\approx 2\Gamma_{gm}/\Gamma_{m}, \quad x_2>>x_1.
\end{equation}
In purely spontaneous relaxation in gases, when
$\Gamma_{ik}\approx (1/2) (\Gamma_{i}+\Gamma_{k})$, expression (7)
turns into
\begin{equation}
x_1=1+\Gamma_{g}/\Gamma_{m},\quad x_1\rightarrow 1,\quad \Gamma_{g}/\Gamma_{m}
\rightarrow 0.
\nonumber
\end{equation}
In the other limiting case $(\gamma_{mn}<<\Gamma_{m})$ and for purely spontaneous
relaxation in gases we have
\begin{eqnarray}
&x_1=1+\dfrac{\Gamma_{n}\Gamma_{g}}{\Gamma_{m}(\Gamma_{m}+\Gamma_{n}+
\Gamma_{g})};\quad x_2=1+\dfrac{\Gamma_{n}}{\Gamma_{m}};&\nonumber\\
&x_3=1+\dfrac{\Gamma_{n}(2\Gamma_{m}+\Gamma_{g})}
{\Gamma_{m}(\Gamma_{m}+\Gamma_{g}-\Gamma_{n})}\quad
(\Gamma_{m}+\Gamma_{g}> \Gamma_{n}).&\nonumber
\end{eqnarray}
Thus the most favorable relationship in this is
$\Gamma_{n}>>\Gamma_{m}>> \Gamma_{g}$. At the same time,
interference process are effective everywhere to the right of the
curve $\varkappa_1(x)$, and a slight excess of level $g$
population over level $n$ population is sufficient {\it for
amplification to take place in a strong field $E$ even if
$n_g-\rho_{nn} (\varkappa)< 0$}.

If $n$ is the ground level, it is necessary to pass to the limits as follows
\begin{equation}
\Gamma_{n}\rightarrow 0,\quad \Gamma_{m}-\gamma_{mn}\rightarrow 0,
\quad (\Gamma_{m}-\gamma_{mn})/\Gamma_{n}\rightarrow 1.
\nonumber
\end{equation}
From (6) we have
\begin{equation}
x_1=1+\dfrac{2\Gamma_{gm}-\Gamma_{m}}{2\Gamma_{gm}+\Gamma_{m}},\quad
x_2=2.\nonumber \tag{6$'$}
\end{equation}
Thus, in solids the value of $x_1$ in the given case is nearly two and
differs little from the value of $x_2$.

For gas and purely spontaneous relaxation (6') gives $x_1>1$ if $\Gamma_{m}>>
\Gamma_{g}$. In this case with $n_m=0$ a slight additional excitation of
the $g$ level is sufficient for the strong field ($\varkappa >> 1$) to
produce amplification.

From the above discussion follows that the contribution of nonlinear
interference processes to amplification is most significant under the
following conditions. In the considered transition scheme, the common level
should be the broadest, the final level of the two-photon transition should
be broader than the starting level, and the relaxation of the upper
strong-field transition level should take place mainly by decay onto the
lower level of this transition.

Let us evaluate the effect of nonlinear interference processes in the
interaction of the transitions $3s_2-2p_4$ and $2s_2-2p_4$ in neon. The
following relaxation constants are used $\Gamma_{m}= 3\cdot 10^7 {\rm sec}^
{-1}$, $\Gamma_{n}= 5\cdot 10^7{\rm sec}^{-1}$, $\Gamma_{g}= 10^7 {\rm sec}^
{-1}$, $\gamma_{mn}=\gamma_{gn}=0.5\cdot 10^7 {\rm sec}^{-1}$. Thus,
$x_1\approx 1.2$ and $x_2\approx 3$ if the strong-field transition is $3s_2-
2p$, and $x_1\approx 3.1$ and $x_2\approx 11$ otherwise. In the first case
$(x_2 -x_1)/x_1\approx 150\%$, while in the second
$(x_2 -x_1)/x_1\approx 255\%$. In both cases (5) is satisfied as an inequality.

Let us now analyze the effect of the strong field, given by the
common denominator in (2), on the amplification factor at the
center of the line. This term reflects the splitting of energy
levels $m$ and $n$ under the effect of the strong field. Since
$|G|^2$ enters both the numerator and denominator of (2) with
different weights, the absolute value of the factor $\alpha_\mu$
at the line center first increases and then falls with increasing
intensity of the external field $\varkappa>\varkappa_1$ for fixed
$x > x_1$.

Thus, for any $x > x_i$ there is an optimum external field at
which the factor $\alpha_\mu$ at the line center is maximum. The
optimum field $\varkappa_{opt}$ as a function of the ratio $x =
\Delta n_{mn}/\Delta n_{gn}$ is given by
\begin{equation}
\varkappa_{opt}(x)=\varkappa_{1}(x)\{1+\sqrt{1+[x_1\varkappa_{1}(x)]^{-1}
(2\tau^2\Gamma_{gm}\Gamma_{gn}x+x_1)}\},\quad x>x_1.
\end{equation}
Here $x_1$ and $\varkappa_1(x)$ are given by (6). With increasing
$x$ ($x > x_1$) the optimum field decreases at the limit as
$x_1/x\{1+\sqrt{1+2\tau^2\Gamma_{gm}\Gamma_{gn}}\}$. For any fixed
field, $\alpha_{\mu}$ depends linearly on $x$. The dependence of
$\alpha_\mu(0)$ on $\varkappa$ and $x$ is qualitatively
illustrated in Fig. 3. For example, for the neon transitions
considered above (strong-field transition $3s_2-2p_4$) $x =4. 14$
corresponds to an optimum field $\varkappa_{opt}=2$. Estimates
indicate that for these optimum values of $\varkappa$ and $x$, the
change in $\alpha_{\mu}$ accompanied by sign inversion is quite
large $\alpha_\mu/\alpha_\mu^0=-32$, and the optimum is sharply
pronounced. Thus, for $\varkappa$ equal to one half of its optimum
value the absolute value of $\alpha_\mu$ decreases by a factor of
nearly three. For $\varkappa$ $50\%$ greater than
$\varkappa_{opt}$, $|\alpha_\mu|$ drops by a factor of $60$.

In conclusion we wish to thank S. G. Rautian for a valuable
discussion.\vspace{1cm}

\centerline{LITERATURE  CITED}\vspace{5mm}

\noindent
1. V. M. Fain, Ya. I. Khanin, and E. G. Yashchin, Izv. Vuzov, Radiofizika, 5, 697 (1962).

\noindent 2. G. E. Notkin, S. G. Rautian, and A. A. Feoktistov,
Zh. Eksper. Teor. Fiz., 52, 1673 (1967).

\noindent
3. T. Ya. Popova, A. K. Popov, S. G. Rautian, and. I. Sokolovskii, Zh. Eksper. Teor. Fiz., 57,
850 (1969).

\noindent
4. H. K. Holt, Phys. Rev. Lett., 19, 1275 (1967); 20, 410 (1968),

\noindent
5. T. Ya. Popova, A. K. Popov, S. Rautian, and. A. Feoktistov, Zh. Eksper. Teor. Fiz., 57,
444 (1969).
                                                                           -
\end{document}